\documentclass[journal]{IEEEtran}
\usepackage{amsmath,amsfonts}
\usepackage{cite}
\usepackage{algorithmic}
\usepackage{array}
\usepackage{acronym} 
\usepackage[caption=false,font=normalsize,labelfont=sf,textfont=sf]{subfig}
\usepackage{textcomp}
\usepackage{stfloats}
\usepackage{tabularx}
\usepackage{xcolor}
\usepackage{url}
\usepackage{verbatim}
\usepackage{multirow}
\usepackage{float}
\usepackage{graphicx}
\def\BibTeX{{\rm B\kern-.05em{\sc i\kern-.025em b}\kern-.08em
T\kern-.1667em\lower.7ex\hbox{E}\kern-.125emX}}
\usepackage{balance}
\usepackage{multirow}
\usepackage{hhline}
\acresetall

\usepackage{fancyhdr}

\begin{document}
\title{A Multi-Objective Learning Approach for Adaptive Waveform Selection in Integrated Sensing and Communications Systems}
\author{Ahmet Yazar, Yusuf İslam Demir, Ahmed Naeem, Seyit Karatepe\\This work has been submitted to the IEEE for possible publication. Copyright may be transferred without notice, after which this version may no longer be accessible.

\thanks{A. Yazar, Y. İ. Demir, and S. Karatepe are with Turk Telekom R\&D, Ankara, 06080, Türkiye. A. Yazar is also with the Department of Electronics Engineering, Gebze Technical University, Kocaeli, 41400, Türkiye. S. Karatepe is also with the Department of Electrical and Electronics Engineering, Istanbul Medipol University, Istanbul, 34810, Türkiye (e-mail: \{ahmet.yazar2, yusufislam.demir, seyit.karatepe\}@turktelekom.com.tr).}

\thanks{A. Naeem is with the School of Electrical Engineering and Computer Science, National University of Sciences and Technology (NUST), Islamabad, 44000, Pakistan (e-mail: ahmed.naeem@seecs.edu.pk).}

\thanks{This work is supported by The Scientific and Technological Research Council of Türkiye (TÜBİTAK) 1515 Frontier R\&D Laboratories Support Program for Türk Telekom 6G R\&D Lab under project number 5249902.}}

\markboth{Journal of \LaTeX\ Class Files,~Vol.~18, No.~9, November~2023}%
{How to Use the IEEEtran \LaTeX \ Templates}

\maketitle

\begin{abstract}
Integrated Sensing and Communications (ISAC) has emerged as a key enabler for sixth-generation (6G) wireless systems by jointly supporting data transmission and environmental awareness within a unified framework. However, communication and sensing functionalities impose inherently conflicting performance requirements, particularly in heterogeneous networks where users may demand sensing-only, communication-only, or joint services. Selecting a waveform that satisfies diverse service demands therefore becomes a challenging multi-objective decision problem. In this paper, a multi-objective learning approach for adaptive waveform selection in ISAC systems is proposed. A simulation-driven evaluation framework is developed to assess multiple waveform candidates across communication, sensing, and joint performance metrics. Instead of enforcing scalar utility aggregation, waveform performance is represented in a multi-dimensional objective space where Pareto-optimal candidates are identified for each scenario. A dataset is generated by varying user demand distributions and channel conditions, and multi-label targets are constructed based on Pareto dominance. Machine learning models are trained to learn the mapping between network conditions and Pareto-optimal waveform sets, enabling fast waveform selection under dynamic network states. Simulation results demonstrate that the proposed framework effectively adapts waveform selection to heterogeneous service requirements while preserving sensing–communication trade-offs, providing a forward-looking perspective for 6G and beyond ISAC deployments.
\end{abstract}



\section{Introduction}

\IEEEPARstart{T}{he} evolution toward sixth-generation (6G) wireless networks is expected to transform communication systems from data-centric infrastructures into intelligent platforms capable of both information exchange and environmental awareness \cite{yazar2020vision}. In this vision, Integrated Sensing and Communications (ISAC) has emerged as a key enabling technology that unifies wireless communication and radar sensing within a shared spectral and hardware framework \cite{liu2022isac}, where a comprehensive framework for dual-functional 6G wireless networks is established. By jointly utilizing spectrum, transceiver chains, and signal processing modules, ISAC systems promise enhanced spectral efficiency, reduced hardware redundancy, and improved situational awareness \cite{wei2023isac}. Such capabilities are considered fundamental for future 6G use cases including autonomous vehicles, smart manufacturing, extended reality, and intelligent transportation systems \cite{etsi_isac_2025}.

Despite its conceptual appeal, ISAC introduces significant physical-layer design challenges. Communication and sensing functionalities impose inherently different, and often conflicting, performance objectives. Communication-oriented metrics such as bit error rate (BER), spectral efficiency, and throughput prioritize reliable and high-rate data transmission, whereas sensing performance depends on metrics including range resolution, velocity resolution, ambiguity function sidelobe levels, and clutter suppression capability \cite{li2025mimoofdm}. These objectives may not be simultaneously optimized by a single waveform due to inherent sensing-communication trade-offs in ISAC systems \cite{keskin2024mimootfs}. In heterogeneous network scenarios where users may require sensing-only, communication-only, or joint sensing–communication services, waveform selection becomes a complex multi-objective decision problem \cite{wang2024moo}.

Although orthogonal frequency division multiplexing (OFDM) has been the dominant multicarrier waveform in contemporary cellular systems and remains a widely adopted baseline in early 6G studies due to its flexibility and spectral efficiency, the waveform design space for fully integrated ISAC operation is still evolving, and a systematic survey of ISAC waveform design approaches and open challenges is presented in \cite{zhou2022isac}. While OFDM provides inherent advantages for communication-centric services, alternative waveform structures have been investigated to improve sensing-communication coexistence \cite{liyanaarachchi2021isac}. Chirp-based waveforms offer favorable sensing characteristics such as lower ambiguity sidelobes \cite{li2025mimoofdm}, whereas delay-Doppler-based waveforms such as orthogonal time frequency space (OTFS) demonstrate strong robustness under high-mobility channel conditions \cite{yuan2024ddisac}. In addition, single-carrier and filtered multicarrier structures further expand the waveform candidate space in terms of PAPR, spectral efficiency, and inter-symbol interference robustness \cite{kebede2022review}. These developments highlight the importance of adaptive waveform selection strategies capable of accommodating diverse ISAC requirements \cite{demir2024waveform}.

Existing research on ISAC waveform design can be broadly categorized into communication-centric approaches, radar-centric approaches, and joint waveform optimization methods; a thorough overview of signal processing techniques covering these three research directions is provided in \cite{zhang2021overview}. Communication-centric studies typically extend multicarrier waveforms to incorporate sensing functionalities, leveraging the existing OFDM infrastructure of cellular systems. Radar-centric works, on the other hand, adapt chirp-based structures for limited data embedding. Joint optimization frameworks attempt to balance sensing and communication metrics using weighted objective functions, and the trade-off between radar SNR and achievable communication rate has been formulated for MIMO-OTFS systems in \cite{keskin2024mimootfs}. Constrained optimization techniques have also been explored to simultaneously satisfy sensing and communication performance bounds \cite{mao2022isac}, with particular attention to millimeter-wave (mmWave) and terahertz (THz) frequency bands.

While different approaches provide valuable insights into waveform co-design, many of them rely on scalarized formulations in which sensing and communication objectives are combined into a single weighted utility function. Such scalarization simplifies optimization but may conceal inherent trade-offs and reduce adaptability when user requirements dynamically change. Moreover, most prior works assume fixed service priorities, which restricts their applicability to scenarios where demand profiles remain static. Homogeneous user demand assumptions further limit the applicability of such designs to realistic 6G heterogeneous network deployments.

Recently, machine learning (ML) techniques have been explored for adaptive resource allocation, modulation selection, and waveform parameter optimization in wireless systems, demonstrating the potential of data-driven approaches for waveform parameter assignment in 6G \cite{yazar2020waveform, sazak2024environment}. Environment-aware ML frameworks have been proposed for waveform management in 6G systems, where supervised learning is used to select suitable waveforms for a given coverage area \cite{demir2024waveform}. Deep learning approaches have also been applied to ISAC-specific waveform optimization tasks \cite{zhang2024intelligent}, while environment-aware supervised learning frameworks have investigated waveform selection under dynamic channel and user demand conditions in 5G uplink communications \cite{hancer2023b}. Despite these efforts, a comprehensive multi-objective learning framework that explicitly preserves Pareto trade-offs among sensing and communication metrics while addressing heterogeneous user demands at the system level remains largely unexplored.

Motivated by these gaps, this paper proposes a multi-objective learning framework for adaptive waveform selection in ISAC systems. Instead of enforcing premature scalarization of sensing and communication objectives, waveform evaluation is formulated in a multi-dimensional performance space where Pareto-optimal candidates are identified for each network scenario. A simulation-driven dataset is constructed by modeling heterogeneous service demands, realistic channel conditions, and multiple waveform candidates. For each scenario, sensing-only, communication-only, and joint performance metrics are computed, and Pareto-based multi-label targets are generated. ML models are then trained to predict suitable waveform candidates under varying network states. Unlike conventional approaches that select a single waveform, the proposed framework predicts a Pareto-optimal set of waveform candidates while explicitly preserving sensing-communication trade-offs.

The main contributions of this paper are summarized as follows:

\begin{itemize}
    \item A simulation-driven evaluation framework is developed to assess multiple waveform candidates under sensing, communication, and joint performance metrics.
    \item A Pareto-dominance based labeling mechanism is proposed to identify waveform candidates that provide balanced trade-offs between sensing and communication objectives.
    \item A large-scale dataset consisting of diverse network scenarios is generated by varying user demand distributions and channel conditions.
    \item ML models are trained to learn the mapping between network state features and Pareto-optimal waveform sets, enabling adaptive waveform selection.
    \item Demand-aware waveform selection maps are constructed to analyze the learned decision regions and to provide insights into waveform sensitivity with respect to channel dynamics and service demand distributions.
\end{itemize}

The remainder of this paper is organized as follows. Section II introduces the system and signal models considered for the ISAC scenario. Section III presents the waveform candidates investigated in this work and summarizes their key sensing and communication characteristics. Section IV describes the proposed multi-objective waveform evaluation framework. Section V details the simulation-based dataset generation methodology and the ML models used for adaptive waveform selection. Section VI provides the dataset statistics and numerical results, including the evaluation of the learning-based waveform selection framework. Finally, Section VII concludes the paper.


\section{SYSTEM MODEL}

ISAC enables communication service delivery and environmental sensing over a shared wireless infrastructure, thereby making waveform management a fundamental system-level design problem in heterogeneous wireless networks \cite{liu2022isac}. In this section, the network topology, user and coverage assumptions, channel abstractions, and scenario-generation principle adopted throughout the manuscript are defined. The resulting system model is intended to provide a common physical and operational basis for the subsequent waveform comparison and adaptive selection framework.

\subsection{Network Topology, User Model, and Coverage Assumptions}

A single-cell ISAC network is considered, in which a base station (BS) provides downlink communication service over a finite coverage region while simultaneously supporting sensing functionality over the same area. Let $\mathcal{U}$ denote the set of active users in the cell, with cardinality $U = |\mathcal{U}|$. The user set is categorized into three service classes,
\begin{equation} \begin{aligned} \mathcal{U} &= \mathcal{U}_{\mathrm{S}} \cup \mathcal{U}_{\mathrm{C}} \cup \mathcal{U}_{\mathrm{SC}} \end{aligned} \label{eq:user_partition} \end{equation}
where $\mathcal{U}_{\mathrm{S}}$, $\mathcal{U}_{\mathrm{C}}$, and $\mathcal{U}_{\mathrm{SC}}$ denote the sensing-only, communication-only, and joint sensing-and-communication user sets, respectively. This classification reflects the heterogeneous service structure of the cell and forms the basis of the scenario abstraction adopted later. Such heterogeneous user requirements are aligned with the vision of 6G and beyond networks, where users are not restricted to communication-only services.

The waveform adaptation task is formulated at the BS side as a \emph{cell-level} decision problem rather than a per-user waveform assignment problem. Accordingly, during each decision interval, one or more suitable waveform candidates are identified from a predefined candidate pool according to the current cell state. If the $m$th decision interval is denoted by
\begin{equation}
\mathcal{T}_m = [mT_{\mathrm{dec}},\,(m+1)T_{\mathrm{dec}}),
\label{eq:decision_interval}
\end{equation}
where $T_{\mathrm{dec}}$ is the waveform update period, then the waveform candidate set identified for that network state is assumed to remain valid throughout $\mathcal{T}_m$. In practice, $T_{\mathrm{dec}}$ should be selected in accordance with a representative cell-level coherence interval, and preferably should not exceed a conservative coherence time, so that the network state may be regarded as quasi-static within each decision interval.

\begin{figure}
\centering 
\resizebox{0.8\columnwidth}{!}{
\includegraphics{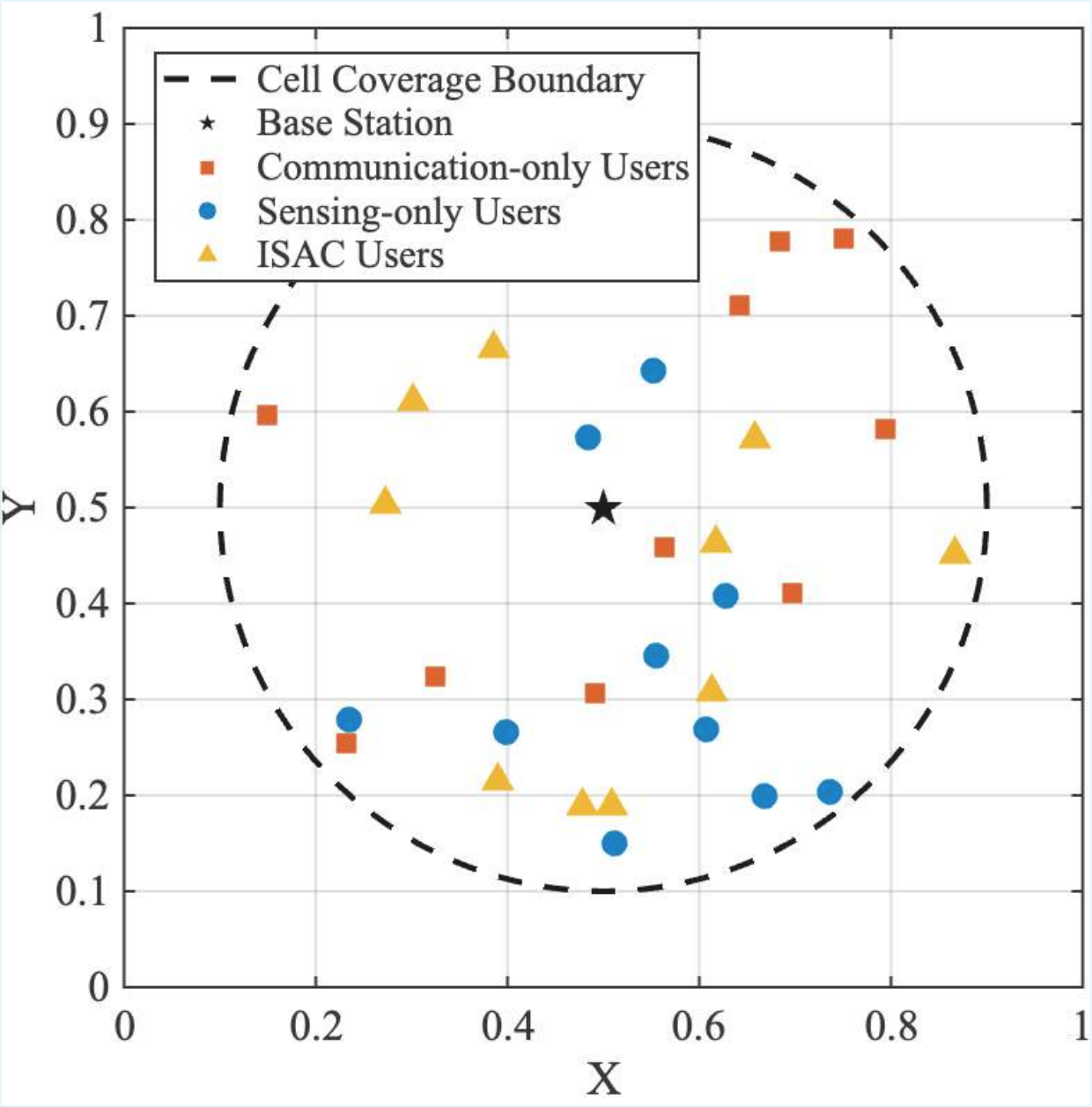}}
\caption{An example of the network coverage layout with 30 users with different requirements.}
\label{Figure25}
\end{figure}

Fig.~\ref{Figure25} illustrates a representative coverage realization containing 30 users with heterogeneous service requirements. The cell boundary defines the spatial support of the considered deployment, while the user markers indicate the coexistence of different service classes within the same region. The figure is intended to visualize the type of heterogeneous network state addressed in this work: users with different requirements may be spatially interleaved across the cell, and the waveform decision is therefore taken according to the aggregate network state observed at the BS rather than being individualized on a user-by-user basis. Hence, the figure should be interpreted as an illustrative realization rather than as a fixed geometric template.

The network state is assumed to be quasi-static within each decision interval. Thus, user locations, large-scale channel conditions, and service requirements are regarded as unchanged over $\mathcal{T}_m$, whereas these quantities may vary from one interval to another due to user mobility, traffic evolution, or environmental changes. Under this assumption, waveform reselection is performed only when a new network state is observed, which establishes a natural bridge between the physical deployment and the later scenario-based decision model.

\subsection{Communication and Sensing Channel Assumptions}

The propagation environment is assumed to be multipath and time-varying, while sensing is assumed to be based on radio echoes returned from surrounding objects and scatterers, as commonly adopted in ISAC system modeling \cite{zhang2021enabling}. In addition to communication-only and sensing-only operation, joint-service users are associated with an effective ISAC channel, in which communication delivery and sensing interaction are coupled. Accordingly, the effective channel seen by an ISAC user does not necessarily reduce to a simple replica of the pure communication-only or sensing-only cases and may exhibit distinct dual-functional behavior \cite{naeem2025novel}. Since the waveform-specific transmit signal structures are introduced separately, the present section focuses only on the channel-level relations needed to connect the physical environment to the later scenario descriptors.

Let $s_k(t)$ denote the baseband-equivalent transmitted signal associated with the selected waveform candidate indexed by $k$. For a communication user $u \in \mathcal{U}$, the received signal is modeled as
\begin{equation}
y_{u,k}(t)
=
\int h_u(t,\tau)\, s_k(t-\tau)\, d\tau + n_u(t),
\label{eq:comm_rx_general}
\end{equation}
where $h_u(t,\tau)$ is the effective time-varying channel response seen by user $u$ and $n_u(t)$ is additive noise. To explicitly reflect multipath propagation and mobility-induced time variation, the user channel may be expressed in delay-Doppler form as
\begin{equation}
h_u(t,\tau)
=
\sum_{\ell=1}^{L_u}
\alpha_{u,\ell}\,
e^{j2\pi \nu_{u,\ell} t}\,
\delta(\tau-\tau_{u,\ell}),
\label{eq:delay_doppler_channel}
\end{equation}
where $L_u$ is the number of significant propagation components, $\alpha_{u,\ell}$ is the complex gain of the $\ell$th path, $\tau_{u,\ell}$ is the associated propagation delay, and $\nu_{u,\ell}$ is the corresponding Doppler shift. This representation makes clear that different users may experience different effective channels due to differences in location, motion, and local scattering conditions.

On the sensing side, the observation at the BS is modeled as the superposition of echoes from surrounding reflectors or targets. The received sensing signal associated with waveform $k$ is written as
\begin{equation}
r_k(t)
=
\sum_{q=1}^{Q}
\beta_q\,
s_k(t-\tau_q)\,
e^{j2\pi f_{D,q} t}
+
n_s(t),
\label{eq:sensing_rx_general}
\end{equation}
where $Q$ denotes the number of dominant reflecting objects, $\beta_q$ is the reflection coefficient of the $q$th object, $\tau_q$ is the round-trip delay, $f_{D,q}$ is the echo Doppler shift, and $n_s(t)$ is sensing noise.

For a joint-service user $u \in \mathcal{U}_{\mathrm{SC}}$, the effective ISAC observation can be represented as
\begin{equation}
\begin{aligned}
z_{u,k}(t) &=
\int h^{\mathrm{I}}_u(t,\tau)\, s_k(t-\tau)\, d\tau \\
&\quad + \sum_{q=1}^{Q_u}
\beta_{u,q}\,
s_k(t-\tau_{u,q})\,
e^{j2\pi f_{D,u,q} t} \\
&\quad + n^{\mathrm{I}}_u(t).
\end{aligned}
\label{eq:isac_rx_general}
\end{equation}
where $h^{\mathrm{I}}_u(t,\tau)$ denotes the communication-oriented component of the effective joint-service channel, the second term captures the sensing-related echo contribution, and $n^{\mathrm{I}}_u(t)$ denotes the effective noise term. Unlike the pure communication-only and sensing-only cases, the ISAC channel jointly reflects data-delivery and environment-sensing interactions and may therefore exhibit a different effective behavior for joint-service users \cite{naeem2025novel}.

Although the communication, sensing, and ISAC channels are not identical, all three are shaped by the same physical scene; therefore, multipath richness, motion, and signal quality influence the corresponding functionalities simultaneously. To convert the detailed propagation relations in \eqref{eq:comm_rx_general}--\eqref{eq:isac_rx_general} into a cell-level waveform selection problem, compact descriptors are adopted. For user $u$, the normalized path powers are defined as
\begin{equation}
p_{u,\ell}
=
\frac{|\alpha_{u,\ell}|^2}{\sum_{i=1}^{L_u} |\alpha_{u,i}|^2},
\qquad
\sum_{\ell=1}^{L_u} p_{u,\ell} = 1.
\label{eq:path_power_norm}
\end{equation}
The mean delay and RMS delay spread of user $u$ are then given by
\begin{equation}
\bar{\tau}_u
=
\sum_{\ell=1}^{L_u} p_{u,\ell}\tau_{u,\ell},
\qquad
\tau_{\mathrm{rms},u}
=
\sqrt{
\sum_{\ell=1}^{L_u}
p_{u,\ell}
(\tau_{u,\ell}-\bar{\tau}_u)^2
},
\label{eq:rms_delay}
\end{equation}
while the mean Doppler and RMS Doppler spread are given by
\begin{equation}
\bar{\nu}_u
=
\sum_{\ell=1}^{L_u} p_{u,\ell}\nu_{u,\ell},
\qquad
\nu_{\mathrm{rms},u}
=
\sqrt{
\sum_{\ell=1}^{L_u}
p_{u,\ell}
(\nu_{u,\ell}-\bar{\nu}_u)^2
}.
\label{eq:rms_doppler}
\end{equation}
Likewise, the effective signal-to-noise ratio of user $u$ is denoted by
\begin{equation}
\gamma_u = \frac{P_{r,u}}{\sigma_u^2},
\label{eq:user_snr}
\end{equation}
where $P_{r,u}$ is the received signal power and $\sigma_u^2$ is the corresponding noise power.

Because waveform selection is performed globally for the cell, the detailed user-dependent quantities are summarized through scenario-level indicators. Specifically, representative cell-level descriptors are formed as
\begin{equation}
\begin{aligned}
\gamma &=
\frac{1}{U}\sum_{u \in \mathcal{U}} \gamma_u, \\
\tau_d &=
\frac{1}{U}\sum_{u \in \mathcal{U}} \tau_{\mathrm{rms},u}, \\
\nu_d &=
\frac{1}{U}\sum_{u \in \mathcal{U}} \nu_{\mathrm{rms},u}.
\end{aligned}
\label{eq:cell_level_descriptors}
\end{equation}
Here, $\gamma$ characterizes the average link quality over the cell, $\tau_d$ captures the effective multipath dispersion, and $\nu_d$ reflects the average mobility-induced channel selectivity. These quantities serve as the principal mathematical bridge between the user-level channel model and the reduced scenario representation used later for waveform evaluation and learning.

In addition to the propagation descriptors above, the operating bandwidth $B$ and the mobility condition of the network are treated as scenario-dependent deployment parameters. The bandwidth determines the available temporal and spectral resolution scale, while mobility governs the severity of time selectivity and the persistence of the network state over each decision interval. Together with $\gamma$, $\tau_d$, and $\nu_d$, these parameters define the operating regime under which candidate waveforms are compared.

\subsection{Scenario Abstraction and Sample Generation Principle}

A scenario is defined as a network snapshot that jointly captures the service composition of the cell, the effective propagation conditions, and the current operating regime. To make this relation explicit, a generic scenario realization may be viewed as
\begin{equation}
\chi
=
\left\{
\Omega,\,
\mathcal{U}_{\mathrm{S}},\,
\mathcal{U}_{\mathrm{C}},\,
\mathcal{U}_{\mathrm{SC}},\,
\gamma,\,
\tau_d,\,
\nu_d,\,
B,\,
\mu
\right\},
\label{eq:scenario_snapshot}
\end{equation}
where $\Omega$ denotes the spatial user layout within the coverage region and $\mu$ denotes the mobility condition associated with the current snapshot. In this form, the scenario description still retains both spatial and service-level information. However, the waveform decision problem is not solved directly from the full geometry; instead, the detailed layout is mapped to an aggregate network state through the induced service composition and the compact channel descriptors in \eqref{eq:cell_level_descriptors}. This establishes the mathematical link between the physical cell realization in Fig.~\ref{Figure25} and the reduced scenario representation adopted later in the manuscript.

The sample generation principle follows directly from \eqref{eq:scenario_snapshot}. Each dataset sample corresponds to one realization of $\chi$, obtained by varying the user mixture, the spatial arrangement, the propagation conditions, and the operating parameters over predefined ranges. For each such realization, all waveform candidates are evaluated under the same scenario conditions. In this manner, differences in sensing, communication, and joint performance can be attributed to the waveform characteristics themselves rather than to inconsistent environmental assumptions. The resulting simulation-driven dataset therefore spans a broad collection of heterogeneous network states and provides a consistent basis for adaptive waveform selection across diverse user-demand and channel regimes.

\section{Considered ISAC Waveforms}

In this section, the considered waveform candidates are briefly introduced under a unified signal representation. A general discrete-time baseband ISAC transmit signal can be expressed as

\begin{equation}
\mathbf{s} = \mathbf{U}\mathbf{x},
\end{equation}
where $\mathbf{x} \in \mathbb{C}^{N}$ denotes the data symbol vector and $\mathbf{U}$ represents the modulation operator defining the waveform structure. Different ISAC waveforms can be interpreted as special cases of $\mathbf{U}$.

Alternatively, in continuous time, a general time-frequency signaling model may be written as

\begin{equation}
s(t) = \sum_{k}\sum_{n} x[k,n] \, g(t-kT) 
e^{j2\pi n\Delta f (t-kT)},
\end{equation}
where $g(t)$ is the transmit pulse, $T$ is the symbol duration, and $\Delta f$ is the subcarrier spacing. The structure of $x[k,n]$ and the transformation applied to it determine the specific waveform realization.

\subsection{OFDM}

In Orthogonal Frequency Division Multiplexing (OFDM), modulation is performed via the inverse discrete Fourier transform (IDFT):

\begin{equation}
\mathbf{s}_{\text{OFDM}} = \mathbf{F}^{H}\mathbf{x},
\end{equation}
where $\mathbf{F}$ is the unitary DFT matrix. In continuous time, the transmitted signal over one symbol duration can be written as

\begin{equation}
s_{\text{OFDM}}(t) = \sum_{n=0}^{N-1} x_n 
e^{j2\pi n \Delta f t}, \quad 0 \le t < T,
\end{equation}
with $\Delta f = 1/T$ ensuring orthogonality.

OFDM offers high spectral efficiency and robustness against multipath fading through cyclic prefix (CP) insertion. Its flexible time-frequency resource allocation is advantageous for communication-centric ISAC scenarios. However, it suffers from high peak-to-average power ratio (PAPR) and range-Doppler ambiguity effects that may limit sensing performance under certain configurations.

\subsection{OCDM}

Orthogonal Chirp Division Multiplexing (OCDM) replaces sinusoidal subcarriers with orthogonal chirp basis functions. Its discrete-time representation can be written as

\begin{equation}
\mathbf{s}_{\text{OCDM}} = \mathbf{F}^{H}\mathbf{C}\mathbf{F}\mathbf{x},
\end{equation}
where $\mathbf{C} = \text{diag}\left(e^{j\pi n^{2}/N}\right)$ is a diagonal chirp matrix.

By employing chirp-based spreading, OCDM exhibits improved robustness in frequency-selective and doubly dispersive channels compared to OFDM. Its structure remains compatible with FFT-based implementations. However, chirp spreading across the full bandwidth may lead to increased out-of-band emissions and similar PAPR characteristics to OFDM.

\subsection{OTFS}

Orthogonal Time Frequency Space (OTFS) modulation maps information symbols into the delay-Doppler domain. Let $\mathbf{X}_{\text{DD}}$ denote the delay-Doppler symbol matrix. The transmitted signal can be expressed as

\begin{equation}
\mathbf{s}_{\text{OTFS}} = \mathbf{F}_{t}^{H}\mathbf{F}_{f}\mathbf{X}_{\text{DD}},
\end{equation}
where $\mathbf{F}_{t}$ and $\mathbf{F}_{f}$ represent Fourier transforms along time and frequency dimensions, respectively.

In continuous form, OTFS can be interpreted as a superposition of time-frequency shifted pulses:

\begin{equation}
s_{\text{OTFS}}(t) = \sum_{n,m} x[n,m] g(t-nT)
e^{j2\pi m\Delta f (t-nT)}.
\end{equation}

Due to its delay-Doppler domain representation, OTFS provides inherent robustness against Doppler shifts and performs well in doubly dispersive channels, making it attractive for high-mobility ISAC scenarios. This comes at the cost of increased computational complexity.

\subsection{FMCW}

Frequency-Modulated Continuous Wave (FMCW) is a chirp-based waveform primarily used for radar sensing. Its baseband representation can be expressed as

\begin{equation}
s_{\text{FMCW}}(t) = e^{j\pi \mu t^{2}},
\end{equation}
where $\mu = B/T$ denotes the chirp rate, $B$ is the sweep bandwidth, and $T$ is the chirp duration.

FMCW offers high range resolution given by

\begin{equation}
\Delta R = \frac{c}{2B},
\end{equation}
where $c$ is the speed of light. While FMCW provides excellent sensing performance and low ambiguity sidelobes under proper design, it does not inherently support high-rate data transmission without additional modulation mechanisms.

\subsection{Single-Carrier}

In single-carrier transmission, the discrete-time signal is directly pulse-shaped:

\begin{equation}
s_{\text{SC}}(t) = \sum_{k} x_k g(t-kT).
\end{equation}

In vector form,

\begin{equation}
\mathbf{s}_{\text{SC}} = \mathbf{x}.
\end{equation}

Single-carrier waveforms exhibit low PAPR, which enhances power amplifier efficiency and nonlinear robustness. Spread-spectrum variants further provide favorable autocorrelation properties beneficial for sensing applications. However, single-carrier schemes are generally more sensitive to severe multipath unless equalization is employed.

The above waveform candidates exhibit distinct sensing and communication characteristics. In this work, their performance is evaluated within a unified multi-objective framework to determine adaptive waveform selection under heterogeneous ISAC service requirements.


\begin{figure}[htp!]
\centering 
\resizebox{0.98\columnwidth}{!}{
\includegraphics{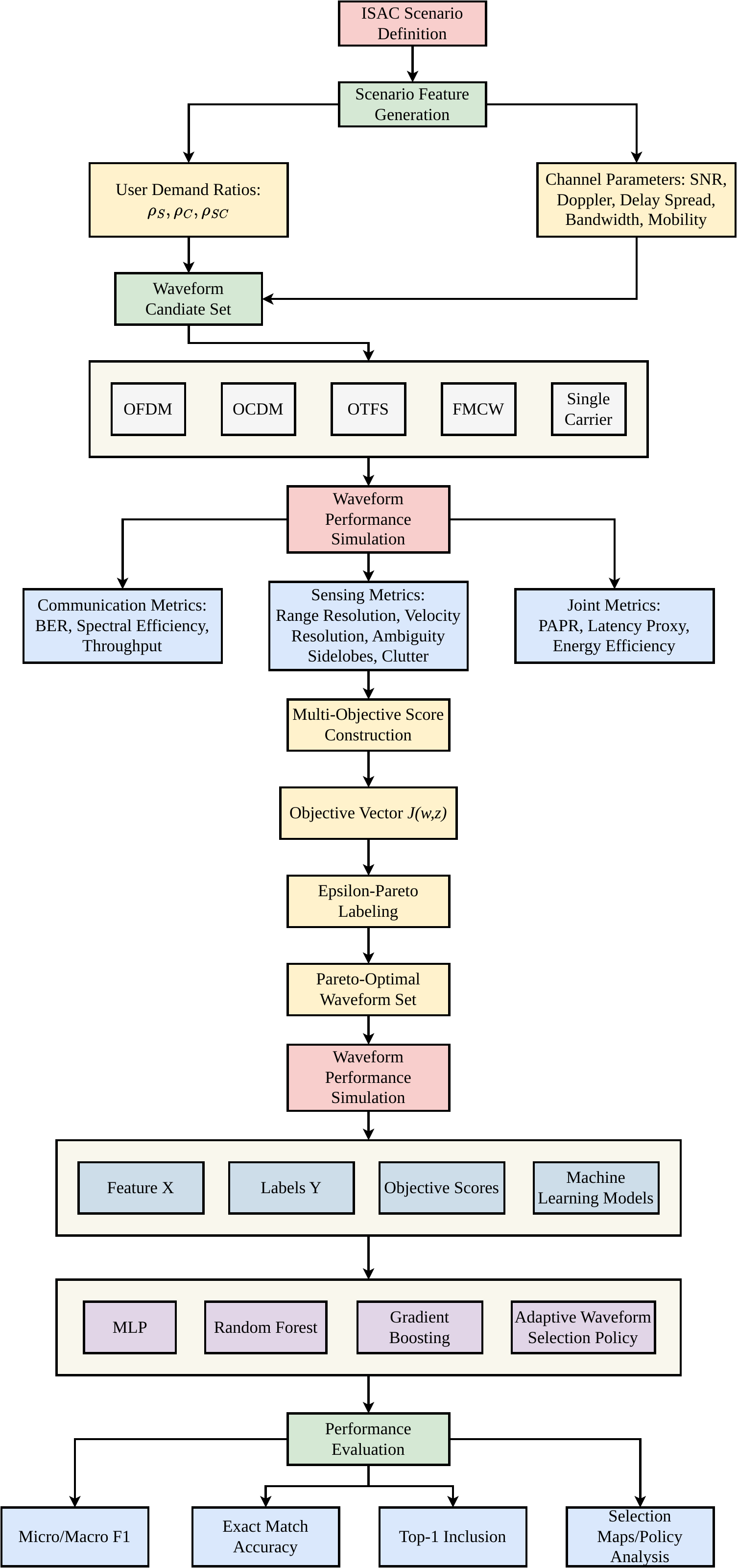}}
\caption{Overview of the proposed multi-objective learning framework for adaptive waveform selection in ISAC systems.}
\label{fig:flowchart}
\end{figure}


\section{Multi-Objective Waveform Evaluation Framework}

The overall framework of the proposed learning-based waveform selection approach is illustrated in Fig.~\ref{fig:flowchart}. The pipeline includes scenario generation, waveform performance simulation, multi-objective evaluation, Pareto-based labeling, and ML-based waveform recommendation.

In heterogeneous ISAC deployments, waveform selection must account for diverse service demands and varying channel conditions. In this work, waveform selection is formulated as a multi-objective decision problem in which Pareto-optimal waveform sets are identified by joinyly evaluating the multiple sensing and communication performance metrics.

\subsection{System Scenario Representation}

Let $\mathbf{z} \in \mathbb{R}^{d}$ denote the scenario feature vector representing the network state, defined as

\begin{equation}
\label{eq:scenario_feature}
\mathbf{z} = 
\left[
\rho_S,\;
\rho_C,\;
\rho_{SC},\;
\gamma,\;
\tau_d,\;
\nu_d,\;
B,\;
\mu
\right],
\end{equation}
where $\rho_S$, $\rho_C$, and $\rho_{SC}$ denote the proportions of sensing-only, communication-only, and joint-service users, respectively; $\gamma$ represents the average SNR; $\tau_d$ is the delay spread; $\nu_d$ is the Doppler spread; $B$ is the bandwidth; and $\mu$ characterizes user mobility.

Let $\mathcal{W} = \{w_1, w_2, \dots, w_K\}$ denote the candidate waveform set, where $K=5$ in this study.

\subsection{Performance Metric Definition}

For each waveform $w_k \in \mathcal{W}$ and scenario $\mathbf{z}$, three performance categories are evaluated:

\begin{itemize}
    \item Sensing-only performance: $J_S(w_k, \mathbf{z})$
    \item Communication-only performance: $J_C(w_k, \mathbf{z})$
    \item Joint performance: $J_J(w_k, \mathbf{z})$
\end{itemize}

\subsubsection{Sensing Objective}

The sensing objective aggregates normalized metrics including range resolution $\Delta R$, velocity resolution $\Delta v$, ambiguity sidelobe level (ASL), and clutter sensitivity. The sensing score is defined as

\begin{equation}
J_S = 
\alpha_1 \frac{1}{\Delta R}
+ \alpha_2 \frac{1}{\Delta v}
+ \alpha_3 \frac{1}{\text{ASL}}
+ \alpha_4 \frac{1}{C_{\text{clutter}}},
\end{equation}
where $\alpha_i$ are weighting coefficients and all terms are normalized.

\subsubsection{Communication Objective}

The communication objective considers BER, spectral efficiency (SE), and throughput $T$:

\begin{equation}
J_C =
\beta_1 (1 - \text{BER})
+ \beta_2 \text{SE}
+ \beta_3 \frac{T}{T_{\max}},
\end{equation}
where $\beta_i$ are normalization weights.

\subsubsection{Joint Objective}

The joint objective accounts for practical system considerations including PAPR, latency $L$, and energy efficiency $\eta$:

\begin{equation}
J_J =
\gamma_1 \frac{1}{\text{PAPR}}
+ \gamma_2 \frac{1}{L}
+ \gamma_3 \eta.
\end{equation}

\subsection{User-Demand-Aware Aggregation}

Since user demands are heterogeneous, objective scores are weighted according to user proportions:

\begin{equation}
\label{eq:objective_scores}
\mathbf{J}(w_k, \mathbf{z}) =
\left[
\rho_S J_S,\;
\rho_C J_C,\;
\rho_{SC} J_J
\right].
\end{equation}

Thus, waveform performance is represented in a three-dimensional objective space.

\subsection{Pareto Optimality}

Instead of combining objectives into a single scalar utility, Pareto dominance is employed.

Waveform $w_a$ is said to dominate waveform $w_b$ under scenario $\mathbf{z}$ if

\begin{equation}
\mathbf{J}(w_a,\mathbf{z}) \succeq \mathbf{J}(w_b,\mathbf{z})
\end{equation}
and at least one objective is strictly greater.

The Pareto-optimal waveform set for scenario $\mathbf{z}$ is defined as

\begin{equation}
\mathcal{W}^*(\mathbf{z}) =
\left\{
w_k \in \mathcal{W}
\mid
\not\exists w_j \in \mathcal{W}
\text{ such that }
w_j \succ w_k
\right\}.
\end{equation}

This formulation preserves inherent trade-offs among sensing, communication, and joint objectives without enforcing premature scalarization.

\subsection{Problem Statement}

Given scenario $\mathbf{z}$, the objective is to identify the Pareto-optimal waveform set $\mathcal{W}^*(\mathbf{z})$. In subsequent sections, this multi-objective formulation is leveraged to construct a dataset in which Pareto-optimal waveforms serve as multi-label targets for ML-based adaptive waveform selection.


\section{Dataset Generation and Machine Learning Methodology}

This section describes the simulation-driven dataset construction process and the ML framework used for adaptive waveform selection. The dataset is generated based on the multi-objective evaluation model introduced in Section IV.

\subsection{Scenario Generation and Feature Construction}

The scenario feature vector $\mathbf{z}$ defined in Eq.~\ref{eq:scenario_feature} is adopted to represent each single-cell ISAC configuration. Each dataset sample corresponds to a distinct realization of $\mathbf{z}$, generated through stochastic sampling within predefined operational ranges.

Specifically, the user demand proportions $\rho_S$, $\rho_C$, and $\rho_{SC}$ are randomly generated under the constraint

\begin{equation}
\rho_S + \rho_C + \rho_{SC} = 1,
\end{equation}
ensuring heterogeneous service distributions across scenarios.

Channel and system parameters are varied within realistic ranges to emulate diverse deployment conditions. The average SNR $\gamma$ is sampled within a predefined interval representing low-to-high link quality regimes. The delay spread $\tau_d$ and Doppler spread $\nu_d$ are generated to reflect frequency-selective and mobility-induced channel variations, respectively. The system bandwidth $B$ is varied to capture different sensing resolution capabilities, while the mobility indicator $\mu$ models user dynamics.

This randomized scenario generation strategy enables the constructed dataset to span a wide range of heterogeneous service demands and channel environments, thereby improving the generalization capability of the trained ML models.

\subsection{Waveform Performance Simulation}

For each generated scenario and each waveform candidate $w_k \in \mathcal{W}$, waveform signals are synthesized and transmitted through a multipath and Doppler channel model.

The following steps are performed:

\begin{enumerate}
    \item Baseband signal generation according to the waveform structure defined in Section III.
    \item Channel application including delay spread and Doppler shift.
    \item Additive white Gaussian noise injection according to the target SNR.
    \item Communication metric computation (e.g., BER, spectral efficiency, throughput). To reflect practical implementation differences among candidate waveforms, waveform-dependent throughput overhead factors are incorporated to account for variations in pilot/guard signaling and processing requirements.
    
    \item Sensing metric computation using matched filtering and ambiguity-based measures (e.g., range resolution, velocity resolution, sidelobe levels).
    \item Joint metric computation including PAPR, latency proxy, and energy efficiency.
\end{enumerate}

The objective vector for each waveform under scenario $\mathbf{z}$ is obtained as in Eq.~\ref{eq:objective_scores}.

All performance metrics are normalized to ensure comparable scaling across objectives.

\subsection{Pareto-Based Multi-Label Construction}

For each generated scenario, waveform objective vectors are computed as described in Section IV. Based on the Pareto dominance principle defined previously, non-dominated waveform candidates are identified within the multi-dimensional objective space.

Specifically, for each scenario realization, all waveform objective vectors are compared pairwise, and dominated candidates are eliminated. The remaining non-dominated waveforms constitute the Pareto-optimal set $\mathcal{W}^*(\mathbf{z})$.

Due to near-ties arising from simulation variability and closely spaced objective values, an $\epsilon$-Pareto dominance criterion is employed. In this relaxed dominance formulation, objective differences smaller than $\epsilon$ are treated as negligible during Pareto set construction. This tolerance-based dominance rule mitigates the impact of minor performance fluctuations and prevents excessively large Pareto sets caused by nearly indistinguishable objective values.

To enable supervised learning, a multi-label encoding strategy is adopted. For each scenario, a binary label vector

\begin{equation}
\mathbf{y} \in \{0,1\}^{K}
\end{equation}

is constructed, where $y_k = 1$ if waveform $w_k$ belongs to the resulting $\epsilon$-Pareto-optimal set and $y_k = 0$ otherwise.

This procedure ensures that inherent trade-offs among sensing, communication, and joint objectives are preserved in the dataset without enforcing scalar utility aggregation. Consequently, the learning model is trained to identify Pareto-consistent waveform candidates under varying network conditions while maintaining robustness against simulation-induced near-ties.

\subsection{Machine Learning Model Design}

The adaptive waveform selection problem is cast as a multi-label classification task:

\begin{equation}
f_{\theta}: \mathbf{z} \rightarrow \mathbf{y},
\end{equation}
where $f_{\theta}$ denotes a parameterized learning model.

In this work, supervised learning models such as:
\begin{itemize}
    \item Multi-layer perceptron (MLP),
    \item Random forest classifier,
    \item Gradient boosting models,
\end{itemize}
are considered for comparison.

Binary cross-entropy loss is used for multi-label training. The dataset is divided into training, validation, and test subsets.

\subsection{Performance Evaluation Metrics}

Model performance is evaluated using both classification-based and system-level metrics:

\begin{itemize}
    \item Multi-label F1-score,
    \item Exact match accuracy,
    \item Hamming loss,
    \item Top-1 inclusion accuracy (whether the highest-confidence prediction lies in the Pareto set),
    \item Utility regret defined as
\end{itemize}

\begin{equation}
\text{Regret} =
\frac{J^*(\mathbf{z}) - J_{\text{model}}(\mathbf{z})}
{J^*(\mathbf{z})},
\end{equation}
where $J^*(\mathbf{z})$ is the maximum achievable scalar utility under oracle selection and $J_{\text{model}}(\mathbf{z})$ is the utility achieved by the waveform selected by the trained model.

The regret metric quantifies the system-level performance degradation relative to ideal selection and provides a practical evaluation of deployment effectiveness.


\section{Simulation Results and Performance Analysis}

\begin{table*}[htp!]
\centering
\caption{Performance comparison of machine learning models}
\label{tab:ml_results}
\begin{tabular}{lcccccc}
\hline
Model & Micro-F1 & Macro-F1 & Exact Match & Hamming Loss & Top-1 Inclusion & Utility Regret \\
\hline
MLP & 0.834 & 0.808 & 0.397 & 0.182 & 0.960 & 0.666 \\
Random Forest & 0.834 & 0.807 & 0.405 & 0.181 & 0.960 & 0.670 \\
Gradient Boosting & \textbf{0.835} & 0.806 & \textbf{0.406} & \textbf{0.179} & \textbf{0.961} & 0.671 \\
\hline
\end{tabular}
\end{table*}

This section presents the dataset characteristics and numerical results obtained using the proposed multi-objective waveform selection framework. The experiments are conducted using the simulation-based dataset generated as described in Section V.

\subsection{Dataset Characteristics}

The generated dataset contains 20,000 scenario realizations with heterogeneous user demand distributions and varying channel conditions. Each scenario is labeled using the resulting $\epsilon$-Pareto-optimal waveform set.

\begin{figure}[t]
\centering
\includegraphics[width=0.75\linewidth]{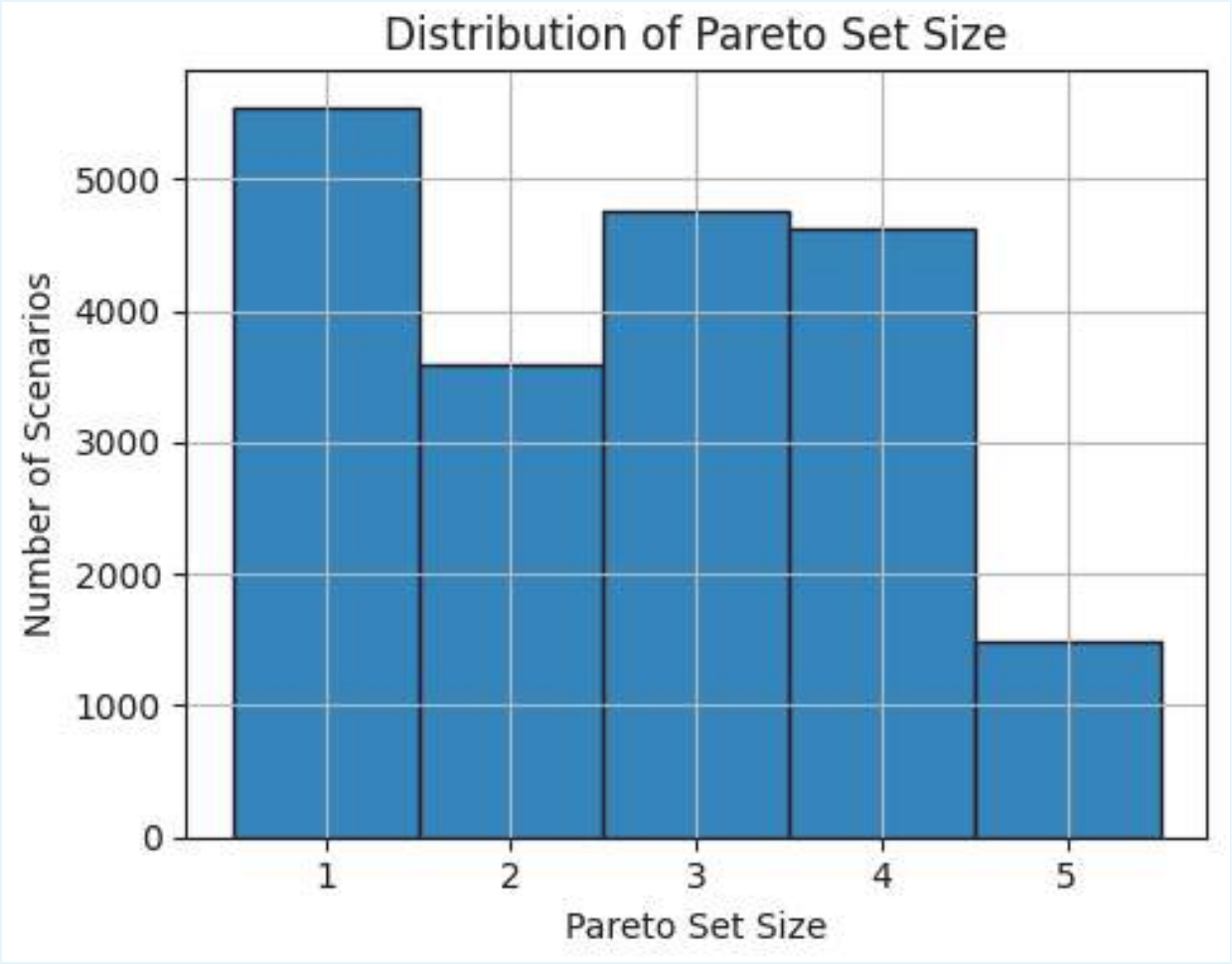}
\caption{Distribution of Pareto set sizes across generated scenarios.}
\label{fig:pareto_size}
\end{figure}

Fig.~\ref{fig:pareto_size} shows the distribution of Pareto set sizes across the generated scenarios. The results indicate that the average Pareto set size is approximately 2.65 with a median value of 3. This confirms that multiple waveform candidates frequently coexist as non-dominated solutions due to inherent sensing-communication trade-offs. In particular, scenarios with three or four Pareto-optimal candidates occur frequently, demonstrating the non-trivial nature of the waveform selection problem.

\begin{figure}[t]
\centering
\includegraphics[width=0.75\linewidth]{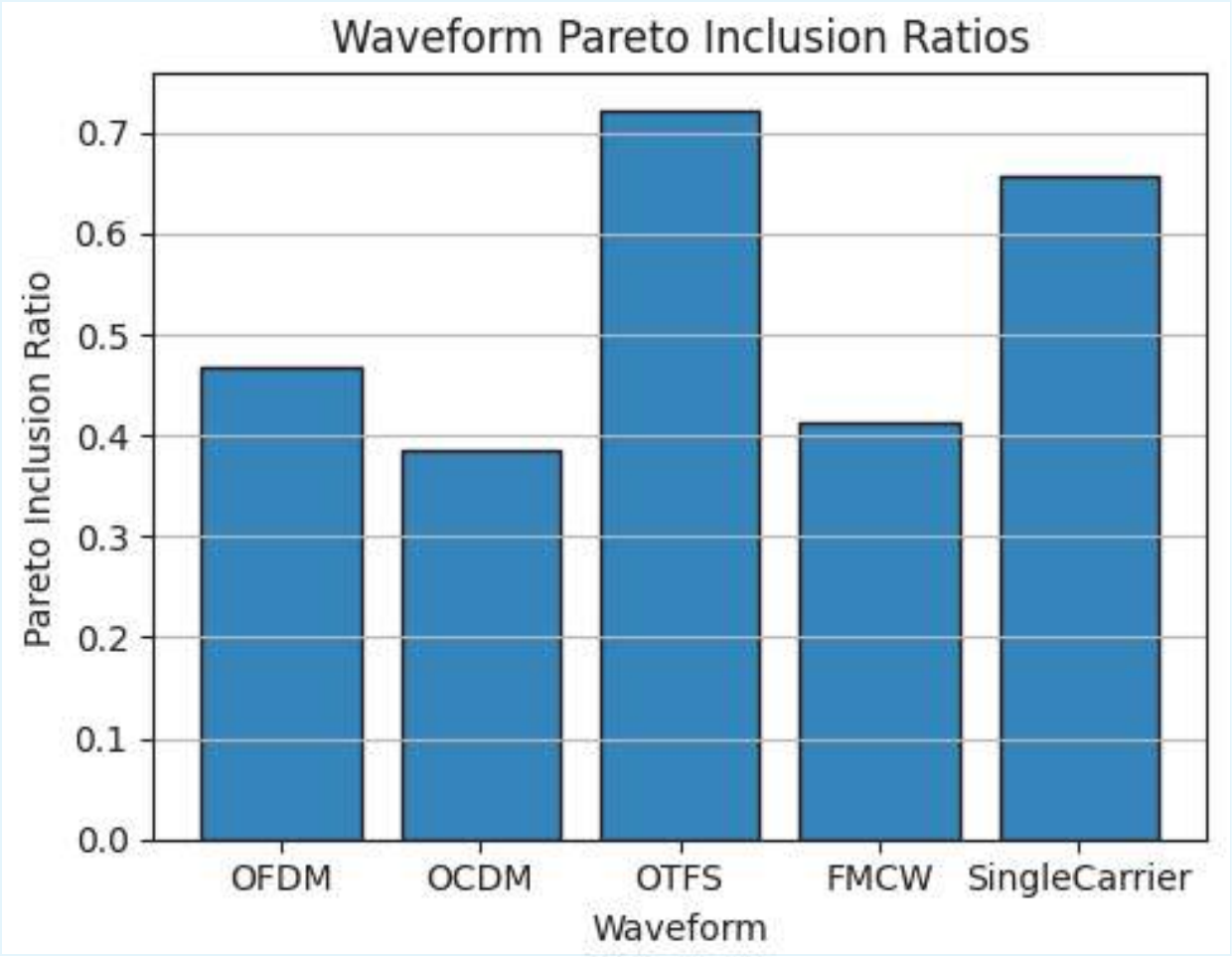}
\caption{Pareto inclusion ratios of waveform candidates across generated scenarios.}
\label{fig:waveform_ratio}
\end{figure}

To further analyze the dataset properties, the Pareto inclusion ratios of the candidate waveforms are computed as the fraction of scenarios in which each waveform appears in the Pareto-optimal set. The resulting ratios are illustrated in Fig.~\ref{fig:waveform_ratio}. The results show that OTFS and the single-carrier waveform appear most frequently in the Pareto set, while OFDM, OCDM, and FMCW also become optimal under certain operating regimes. These statistics confirm that all waveform candidates contribute to the Pareto-optimal solutions under different scenario conditions, ensuring a balanced multi-label learning problem.

\subsection{Learning-Based Waveform Selection Performance}

The supervised learning models described in Section V are trained using the generated dataset. The dataset is divided into training, validation, and test subsets containing 14,000, 3,000, and 3,000 samples, respectively. The validation set is used for model tuning, while the final evaluation is conducted on the held-out test set.

Table~\ref{tab:ml_results} summarizes the performance of the considered machine learning models. The results show that all models achieve strong performance in predicting Pareto-consistent waveform candidates. In particular, the gradient boosting model achieves the highest micro-F1 score and the best exact-match accuracy among the evaluated models.

Furthermore, the top-1 inclusion accuracy exceeds 96\% for all models. This indicates that the waveform with the highest predicted confidence lies within the ground-truth Pareto set with very high probability. Such behavior is particularly desirable in practical deployments where a single waveform must be selected for the entire coverage area.

\subsection{System-Level Decision Quality}

Beyond classification-based metrics, it is also important to evaluate the system-level quality of the waveform selection decisions. In practical deployments, the base station ultimately selects a single waveform for the entire coverage area based on the model predictions. Therefore, the selected waveform should ideally belong to the Pareto-optimal candidate set obtained through the multi-objective evaluation framework.

To assess this aspect, the top-1 inclusion metric is considered. This metric measures the fraction of scenarios in which the waveform with the highest predicted confidence lies within the ground-truth Pareto set. As shown in Table~\ref{tab:ml_results}, all evaluated learning models achieve top-1 inclusion accuracies above 96\%, indicating that the predicted waveform is highly likely to correspond to a Pareto-consistent solution.

These results demonstrate that the proposed learning-based framework is capable of selecting waveform candidates that remain consistent with the underlying sensing-communication trade-offs. Consequently, the framework enables practical waveform selection decisions while avoiding the computational complexity associated with exhaustive multi-objective evaluation for every scenario.

To further understand how waveform preferences vary under different channel conditions and service demand distributions, the next subsection presents demand-aware waveform selection maps across multiple channel regimes.

\subsection{Demand-Aware Waveform Selection Maps}

While the previous results evaluate the statistical properties of the generated dataset and the predictive performance of the learning models, it is also important to visualize how the learned policy behaves across different network conditions. To this end, demand-aware waveform selection maps are generated to illustrate the regions of the service demand space where particular waveform candidates are likely to be selected.

In these visualizations, the horizontal and vertical axes correspond to the proportions of sensing-only and communication-only users, denoted by $\rho_S$ and $\rho_C$, respectively. The proportion of joint-service users is given by $\rho_{SC} = 1-\rho_S-\rho_C$, which restricts the feasible demand region to a triangular domain. For each point in this domain, the trained learning model predicts waveform scores, and the two most probable waveform candidates are identified. A waveform is considered active in a region if it appears among the top-two predicted candidates for that scenario. This top-2 inclusion criterion provides a robust visualization of waveform selection tendencies without forcing a single deterministic decision.

To analyze the influence of different channel characteristics, waveform selection maps are generated under multiple operating regimes. Specifically, two representative values are selected for each key channel parameter: Doppler spread, delay spread, system bandwidth, and SNR. For each regime, five heatmaps are produced corresponding to the candidate waveforms considered in this study.

\begin{figure*}[htp!]
\centering
\includegraphics[width=\linewidth]{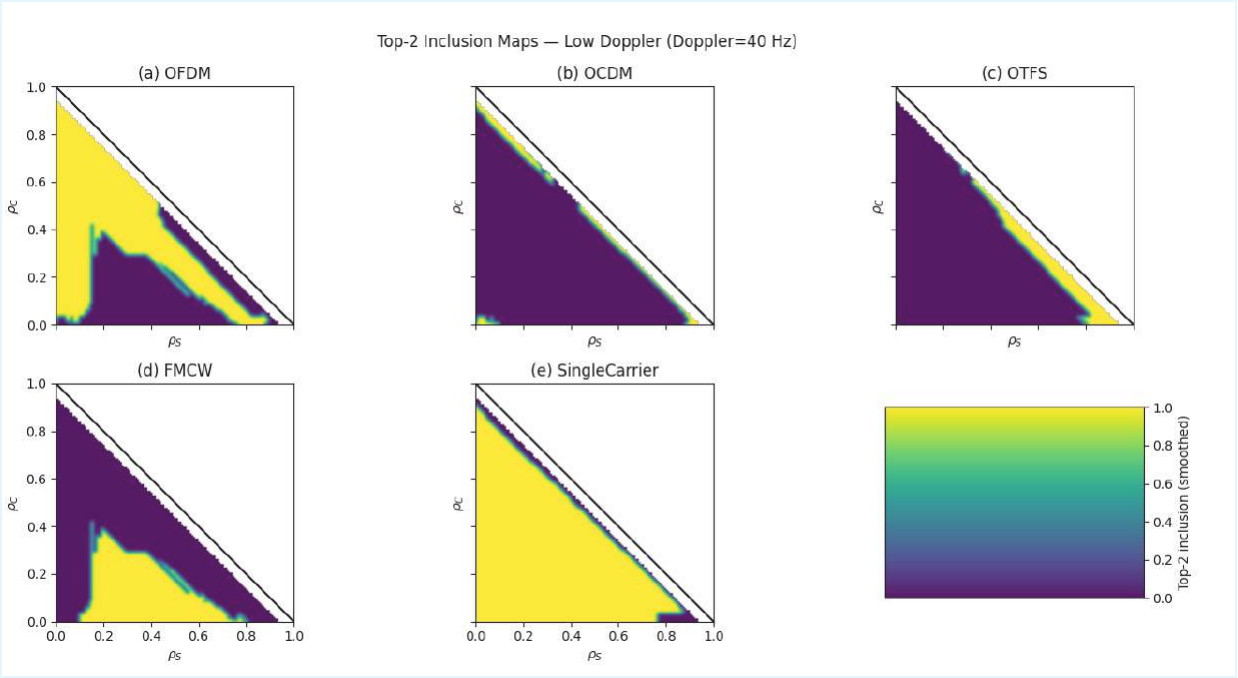}
\caption{Demand-aware waveform selection maps under low Doppler conditions. Each heatmap shows the regions of the service demand space where the corresponding waveform appears among the top-two predicted candidates.}
\label{fig:doppler_maps_low}
\end{figure*}

\begin{figure*}[htp!]
\centering
\includegraphics[width=\linewidth]{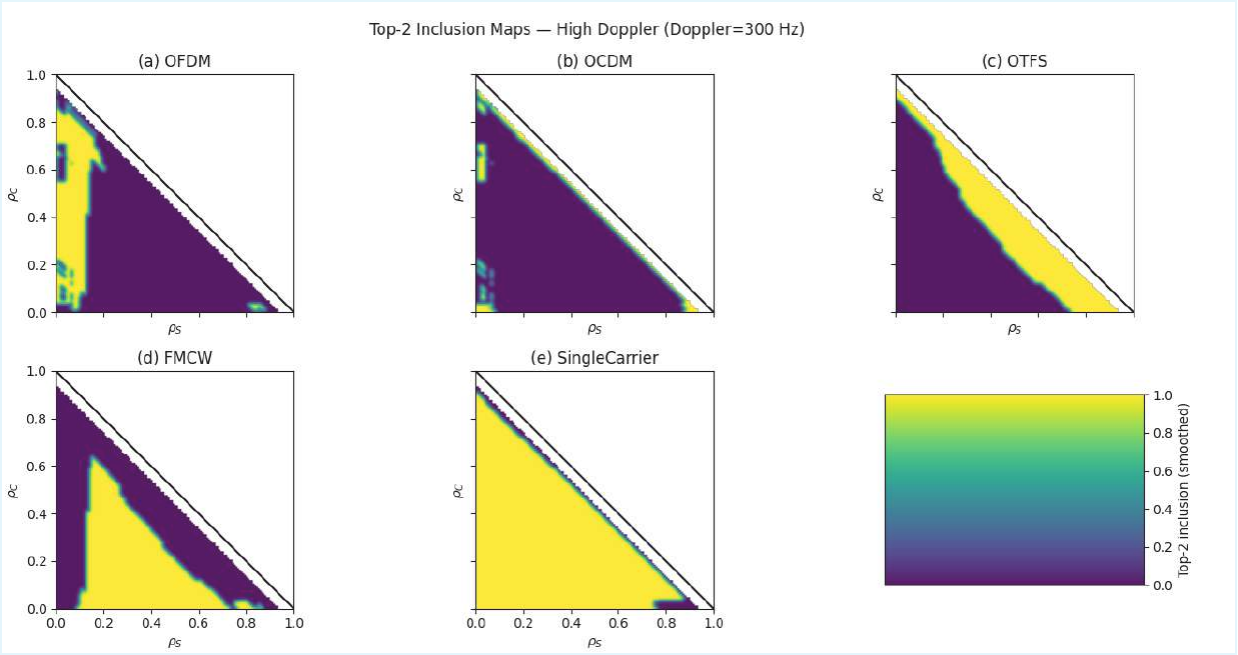}
\caption{Demand-aware waveform selection maps under high Doppler conditions. Increasing Doppler spreads modifies the feasible regions of waveform candidates due to their different robustness to Doppler-induced distortions.}
\label{fig:doppler_maps_high}
\end{figure*}

Figs.~\ref{fig:doppler_maps_low} and~\ref{fig:doppler_maps_high} illustrate the waveform selection patterns under low and high Doppler conditions, respectively. Under low Doppler spreads, several waveform candidates may coexist as viable solutions depending on the service demand distribution. However, as Doppler increases, waveforms that are more robust to Doppler effects tend to dominate larger portions of the demand space.

\begin{figure*}[htp!]
\centering
\includegraphics[width=\linewidth]{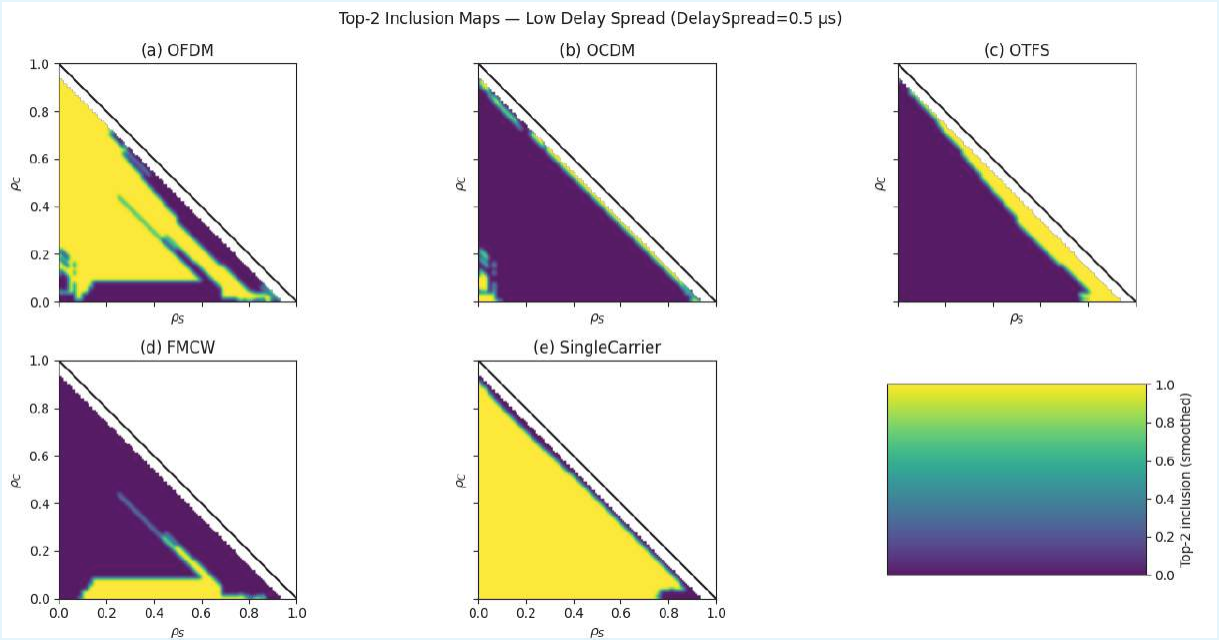}
\caption{Waveform selection maps under low delay spread conditions. In this regime, several waveform candidates remain viable depending on the sensing and communication demand distribution.}
\label{fig:delay_maps_low}
\end{figure*}

\begin{figure*}[htp!]
\centering
\includegraphics[width=\linewidth]{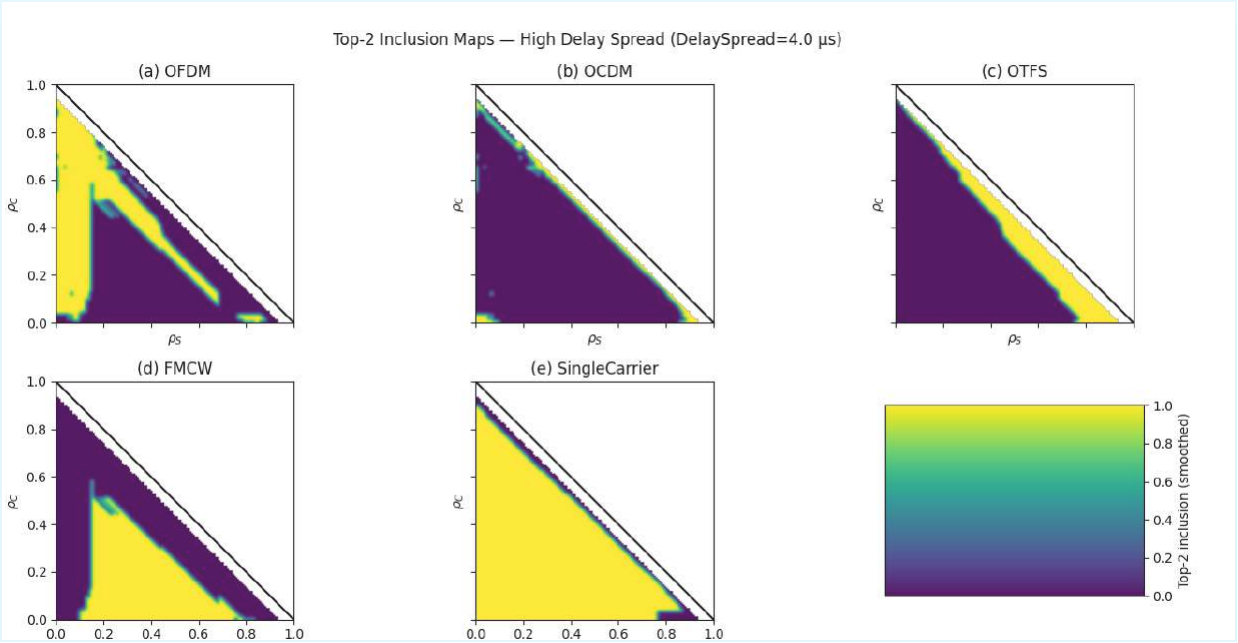}
\caption{Waveform selection maps under high delay spread conditions. Stronger multipath dispersion alters the communication reliability and sensing ambiguity characteristics of the candidate waveforms.}
\label{fig:delay_maps_high}
\end{figure*}

Similarly, Figs.~\ref{fig:delay_maps_low} and~\ref{fig:delay_maps_high} present the selection maps under low and high delay spread conditions. Larger delay spreads introduce stronger multipath dispersion, which alters the communication reliability and sensing ambiguity properties of the waveforms. As a result, the feasible demand regions for certain waveforms expand or shrink depending on their resilience to delay-domain distortions.

\begin{figure*}[htp!]
\centering
\includegraphics[width=\linewidth]{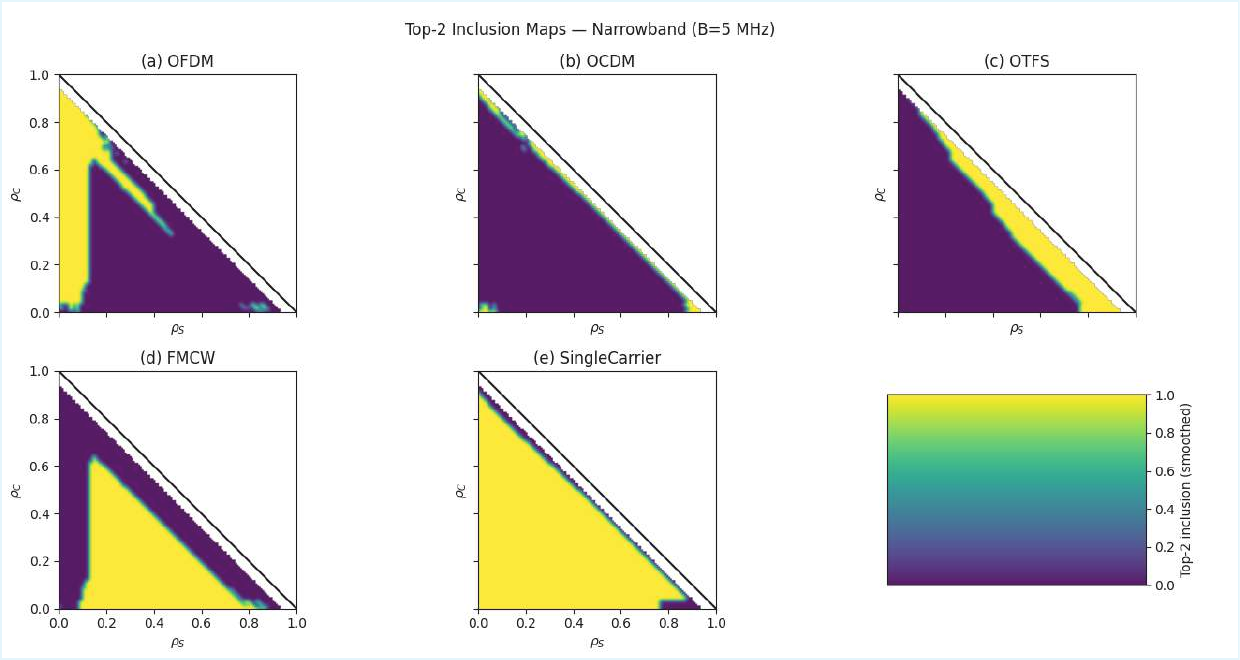}
\caption{Demand-aware waveform selection maps for narrowband operation. Limited bandwidth reduces sensing resolution and communication throughput, affecting waveform preference across the demand space.}
\label{fig:bandwidth_maps_low}
\end{figure*}

\begin{figure*}[htp!]
\centering
\includegraphics[width=\linewidth]{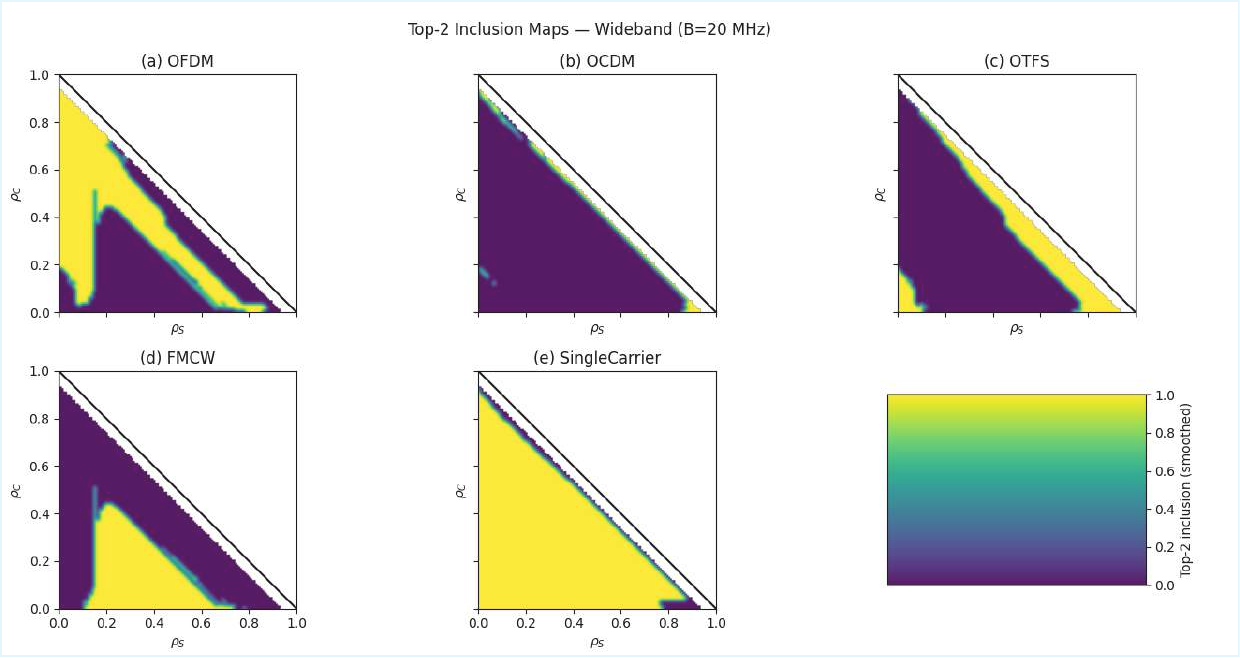}
\caption{Waveform selection maps for wideband operation. Increased bandwidth improves sensing resolution and throughput, leading to expanded feasible regions for several waveform candidates.}
\label{fig:bandwidth_maps_high}
\end{figure*}

The impact of available system bandwidth is illustrated in Figs.~\ref{fig:bandwidth_maps_low} and~\ref{fig:bandwidth_maps_high}. Increasing bandwidth generally improves sensing resolution and communication throughput, which can shift the balance between sensing-oriented and communication-oriented waveform candidates. The resulting maps highlight how the waveform selection policy adapts to different spectral resource conditions.


Overall, these demand-aware waveform selection maps provide an intuitive interpretation of the learned decision policy. The results demonstrate that the proposed framework successfully captures the complex interactions between user service demands and channel conditions, enabling adaptive waveform selection across a wide range of operating regimes. These visualizations also highlight the structured decision regions learned by the model, indicating that waveform preferences emerge from the interplay between service demand composition and channel characteristics rather than arbitrary prediction behavior.

The presented waveform selection maps also provide insight into the sensitivity of waveform preferences to different channel parameters. In particular, variations in Doppler spread, delay spread, bandwidth, and SNR lead to noticeable changes in the decision regions across the demand space. For example, higher Doppler spreads tend to favor waveform structures that exhibit stronger robustness to Doppler-induced distortions, while increased delay spreads influence the communication reliability and sensing ambiguity characteristics of multicarrier and single-carrier waveforms differently. Similarly, wider bandwidths expand the feasible regions of sensing-oriented waveforms due to improved range resolution, whereas higher SNR conditions allow the learning model to exploit waveform structures that provide higher communication efficiency. These observations confirm that the learned policy captures the interaction between channel characteristics and service demand composition, enabling adaptive waveform selection under diverse ISAC operating conditions.


\section{Conclusion}

This paper has presented a multi-objective learning framework for adaptive waveform selection in ISAC systems. Instead of enforcing a scalar utility aggregation from the outset, waveform selection has been formulated as a Pareto-based multi-label decision problem that explicitly preserves the inherent trade-offs among sensing, communication, and joint performance objectives.

A simulation-driven dataset generation methodology has been developed, incorporating heterogeneous user demand profiles, channel variability, and waveform-dependent performance characteristics. Communication metrics (e.g., BER, spectral efficiency, throughput), sensing metrics (e.g., range resolution, velocity resolution, sidelobe levels), and joint metrics (e.g., PAPR, latency proxy, energy efficiency) have been jointly evaluated for multiple waveform candidates. To mitigate near-ties caused by closely spaced objective values, an $\epsilon$-Pareto dominance rule has been employed, ensuring robust and practically meaningful Pareto set construction.

Based on the generated multi-label dataset, supervised learning models have been trained to predict Pareto-consistent waveform candidates under diverse network conditions. The proposed framework enables adaptive waveform recommendation based on predicted Pareto-optimal candidates while maintaining robustness against heterogeneous service demands and varying channel environments. System-level evaluation using utility-regret and Pareto-aware classification metrics demonstrates that the learning-based approach effectively approximates oracle-level decisions without requiring exhaustive real-time optimization.

The presented methodology provides a scalable and extensible foundation for data-driven ISAC waveform adaptation in 6G and beyond wireless systems. Future work may extend this framework toward dynamic online learning, reinforcement learning-based waveform control, and real-time hardware validation in practical ISAC testbeds.

\begin{thebibliography}{10}
\providecommand{\url}[1]{#1}
\csname url@samestyle\endcsname
\providecommand{\newblock}{\relax}
\providecommand{\bibinfo}[2]{#2}
\providecommand{\BIBentrySTDinterwordspacing}{\spaceskip=0pt\relax}
\providecommand{\BIBentryALTinterwordstretchfactor}{4}
\providecommand{\BIBentryALTinterwordspacing}{\spaceskip=\fontdimen2\font plus
\BIBentryALTinterwordstretchfactor\fontdimen3\font minus \fontdimen4\font\relax}
\providecommand{\BIBforeignlanguage}[2]{{%
\expandafter\ifx\csname l@#1\endcsname\relax
\typeout{** WARNING: IEEEtran.bst: No hyphenation pattern has been}%
\typeout{** loaded for the language `#1'. Using the pattern for}%
\typeout{** the default language instead.}%
\else
\language=\csname l@#1\endcsname
\fi
#2}}
\providecommand{\BIBdecl}{\relax}
\BIBdecl

\bibitem{yazar2020vision}
A.~Yazar, S.~Dogan-Tusha, and H.~Arslan, ``{6G Vision: An Ultra-Flexible Perspective},'' \emph{ITU Journal on Future and Evolving Technologies}, vol.~1, no.~1, pp. 121--140, 2020.

\bibitem{liu2022isac}
F.~Liu, Y.~Cui, C.~Masouros, J.~Xu, T.~X. Han, Y.~C. Eldar, and S.~Buzzi, ``{Integrated Sensing and Communications: Toward Dual-Functional Wireless Networks for 6G and Beyond},'' \emph{IEEE Journal on Selected Areas in Communications}, vol.~40, no.~6, pp. 1728--1767, 2022.

\bibitem{wei2023isac}
Z.~Wei, H.~Qu, Y.~Wang, X.~Yuan, H.~Wu, Y.~Du, K.~Han, N.~Zhang, and Z.~Feng, ``{Integrated Sensing and Communication Signals Toward 5G-A and 6G: A Survey},'' \emph{IEEE Internet of Things Journal}, vol.~10, no.~13, pp. 11\,068--11\,092, 2023.

\bibitem{etsi_isac_2025}
{European Telecommunications Standards Institute (ETSI)}, ``{Integrated Sensing and Communications (ISAC); Use Cases and Deployment Scenarios},'' ETSI, Group Report (GR) ISC 001 V1.1.1, Mar. 2025.

\bibitem{li2025mimoofdm}
P.~Li, M.~Li, R.~Liu, Q.~Liu, and A.~Lee~Swindlehurst, ``{MIMO-OFDM ISAC Waveform Design for Range-Doppler Sidelobe Suppression},'' \emph{IEEE Transactions on Wireless Communications}, vol.~24, no.~2, pp. 1001--1015, 2025.

\bibitem{keskin2024mimootfs}
M.~F. Keskin, C.~Marcus, O.~Eriksson, A.~Alvarado, J.~Widmer, and H.~Wymeersch, ``{Integrated Sensing and Communications With MIMO-OTFS: ISI/ICI Exploitation and Delay-Doppler Multiplexing},'' \emph{IEEE Transactions on Wireless Communications}, vol.~23, no.~8, pp. 10\,229--10\,246, 2024.

\bibitem{wang2024moo}
P.~Wang, D.~Han, Y.~Cao, W.~Ni, and D.~Niyato, ``{Multi-Objective Optimization-Based Waveform Design for Multi-User and Multi-Target MIMO-ISAC Systems},'' \emph{IEEE Transactions on Wireless Communications}, vol.~23, no.~10, pp. 15\,339--15\,352, 2024.

\bibitem{zhou2022isac}
W.~Zhou, R.~Zhang, G.~Chen, and W.~Wu, ``{Integrated Sensing and Communication Waveform Design: A Survey},'' \emph{IEEE Open Journal of the Communications Society}, vol.~3, pp. 1930--1949, 2022.

\bibitem{liyanaarachchi2021isac}
S.~D. Liyanaarachchi, T.~Riihonen, C.~B. Barneto, and M.~Valkama, ``{Optimized Waveforms for 5G-6G Communication With Sensing: Theory, Simulations and Experiments},'' \emph{IEEE Transactions on Wireless Communications}, vol.~20, no.~12, pp. 8301--8315, 2021.

\bibitem{yuan2024ddisac}
W.~Yuan, L.~Zhou, S.~K. Dehkordi, S.~Li, P.~Fan, G.~Caire, and H.~V. Poor, ``{From OTFS to DD-ISAC: Integrating Sensing and Communications in the Delay Doppler Domain},'' \emph{IEEE Wireless Communications}, vol.~31, no.~6, pp. 152--160, 2024.

\bibitem{kebede2022review}
T.~Kebede, Y.~Wondie, J.~Steinbrunn, H.~B. Kassa, and K.~T. Kornegay, ``{Multi-Carrier Waveforms and Multiple Access Strategies in Wireless Networks: Performance, Applications, and Challenges},'' \emph{IEEE Access}, vol.~10, pp. 21\,120--21\,140, 2022.

\bibitem{demir2024waveform}
Y.~Ä°slam Demir, A.~Yazar, and H.~Arslan, ``{Waveform Management Approach With Machine Learning for 6G Systems},'' \emph{IEEE Transactions on Network and Service Management}, vol.~21, no.~5, pp. 5432--5444, 2024.

\bibitem{zhang2021overview}
J.~A. Zhang, F.~Liu, C.~Masouros, R.~W. Heath, Z.~Feng, L.~Zheng, and A.~Petropulu, ``{An Overview of Signal Processing Techniques for Joint Communication and Radar Sensing},'' \emph{IEEE Journal of Selected Topics in Signal Processing}, vol.~15, no.~6, pp. 1295--1315, 2021.

\bibitem{mao2022isac}
T.~Mao, J.~Chen, Q.~Wang, C.~Han, Z.~Wang, and G.~K. Karagiannidis, ``{Waveform Design for Joint Sensing and Communications in Millimeter-Wave and Low Terahertz Bands},'' \emph{IEEE Transactions on Communications}, vol.~70, no.~10, pp. 7023--7039, 2022.

\bibitem{yazar2020waveform}
A.~Yazar and H.~Arslan, ``{A Waveform Parameter Assignment Framework for 6G with the Role of Machine Learning},'' \emph{IEEE Open Journal of Vehicular Technology}, vol.~1, no.~1, pp. 156--172, 2020.

\bibitem{sazak2024environment}
H.~Sazak and A.~Yazar, ``{Environment-aware Intelligent Numerology Control Approach for 5G and Beyond Systems},'' \emph{International Journal of Communication Systems}, vol.~37, no.~9, p. e5768, 2024.

\bibitem{zhang2024intelligent}
J.~Zhang, S.~Guo, S.~Gong, C.~Xing, N.~Zhao, D.~W. Kwan~Ng, and D.~Niyato, ``{Intelligent Waveform Design for Integrated Sensing and Communication},'' \emph{IEEE Wireless Communications}, vol.~32, no.~1, pp. 166--173, 2025.

\bibitem{hancer2023b}
A.~{Hançer} and A.~Yazar, ``{Waveform Decision Method with Machine Learning for 5G Uplink Communications},'' \emph{International Journal of Engineering Research and Development}, vol.~15, no.~2, p. 820â€“827, 2023.

\bibitem{zhang2021enabling}
J.~A. Zhang, M.~L. Rahman, K.~Wu, X.~Huang, Y.~J. Guo, S.~Chen, and J.~Yuan, ``{Enabling Joint Communication and Radar Sensing in Mobile Networks - A survey},'' \emph{IEEE Communications Surveys \& Tutorials}, vol.~24, no.~1, pp. 306--345, 2021.

\bibitem{naeem2025novel}
A.~Naeem, E.~M. Amhoud, and H.~Arslan, ``{A Novel User Association Scheme for Joint Radar and Communication in Cell-Free mMIMO Systems},'' \emph{IEEE Transactions on Vehicular Technology}, 2025.

\end{thebibliography}
\end{document}